\documentclass[preprint,nofootinbib]{revtex4}
\usepackage{amssymb}
\usepackage{multirow}
\usepackage{array} 

\usepackage{hyperref}

\usepackage{amssymb}
\usepackage{times,fancyhdr}
\usepackage{amsmath}
\usepackage{amsfonts}
\usepackage{epstopdf}

\begin{document}

%............................ definitions ..............
\newcommand{\be}{\begin{equation}}
\newcommand{\ee}{\end{equation}}
\def\bq{\begin{eqnarray}}
\def\eq{\end{eqnarray}}
%........................................................

\title{\bf Can a  tensorial analogue of gravitational force explain away the galactic rotation curves without dark 
matter?\footnote{{\bfseries{Received an `Honorable Mention' in the 2021 Essay Competition of the Gravity Research Foundation}}}}

\author{Ram Gopal Vishwakarma}

 \address{Unidad Acad$\acute{e}$mica de Matem$\acute{a}$ticas\\
 Universidad Aut$\acute{o}$noma de Zacatecas\\
 C.P. 98068, Zacatecas, ZAC, Mexico\\
Email: vishwa@uaz.edu.mx}

\begin{abstract}
The dark matter problem is one of the most pressing problems in modern physics.
As there is no well-established claim from a direct detection experiment supporting the existence of the illusive dark matter that has been postulated to explain the flat rotation curves of galaxies, and since the whole issue of an alternative theory of gravity remains controversial,
it may be worth to reconsider the familiar ground of general relativity (GR) itself for a possible way out.

It has recently been discovered that a skew-symmetric rank-three tensor
field - the Lanczos tensor field - that generates the Weyl tensor differentially,  provides a proper relativistic analogue of the Newtonian gravitational force. By taking account of its conformal invariance, the Lanczos tensor leads to a modified acceleration law which can explain, within the framework of GR itself, the flat rotation curves of galaxies without the need for any
dark matter whatsoever.

\end{abstract}


\maketitle

Dark matter was hypothesized to explain, in terms of the standard local Newtonian gravity, the non-Keplerian behaviour of the rotational velocities of stars and gaseous components of galaxies. The Newtonian theory gives  the rotational velocity $v$ of a test mass in a circular Keplerian orbit at a distance $r$ from the centre of the galaxy 
as $v=\sqrt{GM(r)/r}$, where $M(r)$ is the galactic mass out to radius $r$ from the centre.
This predicts that $v$ should drop as $r^{-1/2}$. In reality however, $v$ remains more or less constant well beyond the visible galactic disc, after rising sharply outside the nuclear region.

 The most common explanation to this discrepancy is that the galaxies contain far more mass than the directly observed luminous matter. That is, this unseen matter is dark and also non-baryonic (to meet the constraints imposed by the theory of nucleosynthesis), which interacts only gravitationally.
But despite intensive searches for the candidates of non-baryonic dark matter particles, both in direct detection experiments and at the Large Hadron Collider at CERN, no such dark matter candidates have been observed.
A second line of explanation to this discrepancy proposes a modification of the theory of gravitation. In this view, a plethora of modified-gravity theories have been proposed, the MOND (MOdified Newtonian Dynamics) being the most popular one \cite{Mond}, though it is an adhoc and empirically motivated phenomenological hypothesis which lacks a theoretical justification.

There have been problems, in the past also, with the application of Newton's law of gravity to the solar system, which were explained away by Einstein's GR.
In doing so, GR provides, in a weak field, corrections to the corresponding Newtonian equations, which take account of the effects that remain unexplained by the Newtonian gravity.
So, how does GR fare with this problem? It is well-known that the metric tensor provides a relativistic analogue of the Newtonian gravitational potential in the case of a weak field. Then, the relativistic analogue of the Newtonian gravitational force is attributed to the Christoffel symbol, as it is given in terms of the derivatives of the metric tensor.
With this analogy thus, one gets the same Newtonian answer and GR does not fare any better than the Newtonian gravity.

 However, this conventional analogy of the gravitational force, in a covariant theory like GR, does not seem satisfying. While the analogue of its potential is ascribed to a tensor - the metric tensor,  the analogue of force - the Christoffel symbol - is not a tensor. So, do we have a proper tensor that can be ascribed to the Newtonian force and reduces to the derivatives of the metric tensor in a weak field and low velocity limit? Interestingly, it has been shown recently that the Lanczos tensor \cite{Lanczos} - a rank-3 tensor which is a fundamental building block of any metric theory of gravity - fulfills these requirements and thus can be interpreted as a relativistic, tensorial analogue of the Newtonian gravitational force \cite{EPJC}. (A brief account of the Lanczos tensor and a summary of Ref. \cite{EPJC} have been provided in the Appendix. For a detailed introduction to the Lanczos tensor, one may see Refs. \cite{EPJC}, \cite{CQG} and the references therein.)
In the following, we shall see that the tensor leads to a force law which modifies the Newtonian force law by augmenting it with a MOND-like term, together with another term in the weak field limit. These terms have been used in the literature to explain the flat rotation curves of galaxies without postulating the exotic dark matter.

As the Lanczos tensor generates the Weyl tensor of the considered manifold and since the latter is linked with the (free) gravitational field, one may naturally expect that the Lanczos tensor too has something to do with the gravitational field.
In order to exemplify that the Lanczos tensor is a true relativistic analogue of the Newtonian force, and also to use its value to solve the problem of the galactic rotation velocities, let us consider the Schwarzschild line element
\be
ds^2=\left(1-\frac{2GM}{c^2 r}\right) c^2 dt^2-\left(1-\frac{2GM}{c^2 r}\right)^{-1} dr^2 -r^2d\theta^2-r^2\sin^2\theta ~d\phi^2.\label{eq:sch}
\ee
It may be mentioned that a static, spherically symmetric line element is invariably used to model the spacetime structure of a typical galaxy while trying to solve the above-mentioned problem in a relativistic theory (see, for example, \cite{Mannheim-Kazanas,Moffat}). 
A typical galactic component, whose gravitational source  is restricted to the interior region of the galaxy, moves, to a reasonable approximation, in vacuum. So, the static spherically symmetric exterior Schwarzschild solution is valid to a reasonable approximation. 

It has not so far been possible to find an algorithm that can produce the Lanczos tensor in any general spacetime. Nor it is possible to integrate the defining equation (\ref{eq:Weyl-Lanczos}) to obtain this tensor in a general case, given the Weyl tensor. Nevertheless, the Lanczos tensor has indeed been
found explicitly in certain special cases, Schwarzschild spacetime being one of them. In this connection,
Novello and Velloso \cite{N-V} have found that if a
unit time-like vector field $V^\alpha= dx^\alpha/ds$ tangential to the trajectory of an observer in a given spacetime is irrotational and shear-free, the Lanczos tensor of the sapcetime is given by
\be
L_{\mu\nu\sigma}=V_{\mu;\kappa}V^\kappa V_\nu V_\sigma - V_{\nu;\kappa}V^\kappa V_\mu V_\sigma.\label{eq:NV}
\ee
By considering $V^\alpha=\delta^\alpha_t (1-\frac{2GM}{c^2 r})^{-1/2}=\left((1-\frac{2GM}{c^2 r})^{-1/2},0,0,0\right)$, which comes out to be irrotational and shear-free in the spacetime (\ref{eq:sch}), the formula (\ref{eq:NV}) provides the Lancozs tensor of this spacetime with only one non-vanishing, independent component:
\be
L_{trt}=\frac{GM}{c^2 r^2},\label{eq:Lan-sch1}
\ee
which is nothing but the gravitational force (given in the units of $c^2$)  on a unit test mass (in Newtonian terminology)   located at radius $r$, moving under the influence of a static, spherically
symmetric source mass $M$ located inside its orbit. Interestingly, the acceleration vector $a^\mu=V^\mu_{~~;\kappa}V^\kappa$ derived for the considered observer $V^\mu$, also seems consistent with this interpretation: $a^\mu$ has only one non-vanishing component which is radial giving $a^r=GM/(c^2 r^2)$. Let us note that this observer is stationary in the considered spacetime and hence the measured acceleration results from the intrinsic geometry of the  spacetime and not from the observer's motion. This justifies our interpreting the tensor $L_{\mu\nu\sigma}$ as the Newtonian gravitational acceleration of a test mass. 
However, if this interpretation is correct,  the tensor $L_{\mu\nu\sigma}$ must reduce to some expression in terms of the gradient of the metric tensor $g_{\mu\nu}$ in a weak field.  We should expect this because it is well-established that $g_{\mu\nu}$ reduces to the Newtonian gravitational potential in a weak field.
This requirement is indeed met as we note in the following.

We have mentioned earlier that  unlike the Weyl tensor, it has not so far been possible to obtain an expression for the Lanczos tensor in terms of the metric tensor in a general case. However, this expression does exist in the case of a weak gravitational field, wherein
the spacetime metric differs minutely from the Minkowskian metric $\eta_{\mu\nu}$, i.e.,
$
g_{\mu\nu}= \eta_{\mu\nu}+ h_{\mu\nu}, ~{\rm where} ~~h_{\mu\nu}^2<<|h_{\mu\nu}|.\label{eq:weak}
$
In this case, the Lanczos tensor can be written as \cite{Lanczos}
\be
L_{\mu\nu\sigma}=\frac{1}{4}\left(h_{\mu\sigma,\nu} - h_{\nu\sigma,\mu} +\frac{1}{6} h_{,\mu}\eta_{\nu\sigma} - \frac{1}{6} h_{,\nu}\eta_{\mu\sigma}\right),  ~~~ {\rm with}~~~ h\equiv h_{\mu\nu}\eta^{\mu\nu},  \label{eq:weakL}
\ee 
 to linear order in the metric perturbations $h_{\mu\nu}$.
It has been shown in \cite{EPJC} that this definition yields a value for $L_{\mu\nu\sigma}$ consistent with (\ref{eq:Lan-sch1}) in a weak field limit of the line element (\ref{eq:sch}).
Having established that the tensor $L_{\mu\nu\sigma}$ presents a proper relativistic analogue of the Newtonian gravitational force\footnote{A word of caution would be in order about the denomination `Lanczos potential' of the Lanczos tensor commonly used in the literature. As has been exemplified in the Appendix, the Lanczos tensor appears as the potential to the Weyl tensor since the former generates the latter differentially in the same way as the electromagnetic potential generates the Faraday tensor. It should however be noted that the Lanczos `potential' does not have the dimensions of the potential energy, in the same way as the Weyl tensor does not have the dimensions of force. It is rather the Lanczos tensor that has the dimensions of force (per unit mass, in the units of $c^2$). Thus the denomination `Lanczos potential' is something of a misnomer and refers to just the electromagnetic analogy.
\\    $~~~~~~$ It may also be mentioned that the interpretation of the Lanczos tensor as an analogue of Newtonian gravitational force, may seem contradictory with the admissibility of the Lanczos tensor by the flat Minkowski spacetime \cite{CQG}. However, it has also been discovered that  some additional constraints/assumptions, besides (\ref{eq:syma}), (\ref{eq:symb}) and (\ref{eq:conditiona}), must be imposed on the tensor to eliminate the surplus degrees of freedom present in it \cite{EPJC}. It is possible that in the presence of such conditions, any non-trivial value of the Lanczos tensor does not exist in the Minkowski spacetime.}, 
let us use it to the problem of the galactic rotation curves.  But wait a moment! If the tensor reproduces just the Newtonian term, how can it solve the problem of the non-Newtonian rotation velocities? Let us however note that this is not the end of the story of Lanczos tensor.

It can be checked that by virtue of the two symmetries given by (\ref{eq:constraints}), the tensor  $ L_{\mu\nu\sigma}$  readily reproduces, through the Weyl-Lanczos equation (\ref{eq:Weyl-Lanczos}), all the algebraic symmetries of the Weyl tensor: $C_{\mu\nu\sigma\rho}=C_{[\mu\nu][\sigma\rho]}$,  $C_{\mu\nu\sigma\rho}=C_{\sigma\rho\mu\nu}$,  $C_{\mu[\nu\sigma\rho]}=0$, $C^{\kappa}_{~~\mu\nu\kappa}=0$. But just these symmetries do not uniquely single out the Weyl tensor, as these are also shared by the Riemann tensor. The Weyl tensorr is augmented with an additional, characteristic symmetry which is analytical, not algebraic - the local conformal invariance:  $\tilde{C}^\mu_{~~\nu\sigma\rho}=C^\mu_{~~\nu\sigma\rho}$ under the transformation $g_{\mu \nu} \rightarrow \tilde{g}_{\mu \nu}=\Omega^2 g_{\mu \nu}$, with the conformal factor $\Omega(x^\alpha)$ being a positive, smooth function of the spacetime coordinates.
 It could not be realized, until recently \cite{EPJC}, that the two defining symmetries of $ L_{\mu\nu\sigma}$ given by (\ref{eq:constraints}) are not enough to incorporate the conformal invariance of Weyl tensor to equation (\ref{eq:Weyl-Lanczos}). 
However, the Lanczos tensor, as a generator of the Weyl tensor through the Weyl-Lanczos generating equation (\ref{eq:Weyl-Lanczos}), must reproduce this symmetry as well.
It has been shown in Ref. \cite{EPJC} that the conformal invariance of Weyl tensor, imposes an additional constraint on $ L_{\mu\nu\sigma}$ besides others - that the tensor must satisfy the condition (\ref{eq:conditiona}) of the trace-freeness, i.e.
$
L^{~~\kappa}_{\mu~~\kappa}=0.
$

As has been explained in the Appendix, this constraint on $ L_{\mu\nu\sigma}$ has already been in use in the literature since the time Lanczos introduced it. But it is used as an optional gauge condition, just in order to reduce the number of independent components of the tensor, and it was not realized that the constraint is in fact a defining symmetry of the Lanczos tensor.
Interestingly with symmetry (\ref{eq:conditiona}), the Lanczos tensor becomes conformally invariant:  $\tilde{L}^\mu_{~~\nu\sigma}=L^\mu_{~~\nu\sigma}$, akin to $\tilde{C}^\mu_{~~\nu\sigma\rho}=C^\mu_{~~\nu\sigma\rho}$ under the transformation $g_{\mu \nu} \rightarrow \tilde{g}_{\mu \nu}=\Omega^2 g_{\mu \nu}$ \cite{EPJC}.

In order to obtain the Lanczos tensor in a trace-free form, we can use the the gauge transformation (\ref{eq:gaugeU}).
By considering $X_\mu=-L_{\mu~~\kappa}^{~~\kappa}/3$, the transformation (\ref{eq:gaugeU}) provides a trace-free form of the tensor by splitting (\ref{eq:Lan-sch1}) into the following three independent contributions
\be\label{eq:Lan-sch}
\left.\begin{aligned}
&\bar{L}_{trt}=\frac{2GM}{3c^2r^2},\\
&\bar{L}_{\theta r \theta}=\frac{GM}{3c^2\left(1-\frac{2GM}{c^2r}\right)},\\
&\bar{L}_{\phi r \phi}=\frac{GM\sin^2\theta}{3c^2\left(1-\frac{2GM}{c^2r}\right)},
 \end{aligned}
 \right\}
\ee
all of which now contribute to the net force applied on the test mass exerted by a static, spherically symmetric system. Let us note that the orbit of the test mass is confined to the equatorial plane $\theta=\pi/2$ in GR (as it is so in the Newtonian theory) \cite{ABS}. This reduces the three components of $ L_{\mu\nu\sigma}$ described in (\ref{eq:Lan-sch}) to only two independent contributions, which seems reasonable.
A tensor constituting an analogue of force in a relativistic metric theory like GR must have two parts,  in the same way as is the case in the relativistic theory of electrodynamics. The first part corresponds to the Newtonian force law, whereas the second part does not have any classical analogue and constitutes an additional, relativistic contribution. (A detailed account thereof will be presented elsewhere shortly.) 
Hence, the net gravitational acceleration (in conventional units) of the test mass  due to the source mass $M(r)$ confined inside the radius $r$, can be approximated  in a weak field $\left(\left[1-\frac{2GM}{c^2r}\right]^{-1}\approx 1+\frac{2GM}{c^2r}\right)$ by
\be
a(r)= \frac{2GM(r)}{3r^2}  +\alpha GM(r)\left(\frac{2GM(r)}{c^2 r}+ 1 \right).\label{eq:acc}
\ee
An adjustable dimensional constant $\alpha$  has been introduced in order to dimensionally homogenize the last two components in equation  (\ref{eq:Lan-sch}), that differ in dimension from the first component. This happens due to the use of the dimensionally nonuniform coordinate system $(ct, r, \theta, \phi)$, wherein the last two coordinates do not have the dimension of the first two.
The rotational velocity $v(r)$ of the test mass is then obtained from $v^2(r)/r= a(r)$ giving
\be
v^2(r)= \frac{2GM(r)}{3r}  +\alpha GM(r)\left(\frac{2GM(r)}{c^2 }+ r \right).\label{eq:vsquare}
\ee
The first term in equations  (\ref{eq:acc}) and (\ref{eq:vsquare}) corresponds to the Newtonian force, which dominates over the rest two terms at small $r$. The second term corresponds to the phenomenological force law proposed by MOND theory, and becomes competitive with the Newtonian term at medium $r$.  At large enough $r$, it is the last term that dominates, thereby providing the desired departure from the standard Newtonian gravity on large distance scales. 
It may be noted that this last term, which corresponds to a linear potential, successfully explains the observed galactic rotation curves without the need for any dark matter, as has been shown in \cite{Mannheim-Kazanas}.

It would be worthwhile to make a few remarks on the conformal Lanczos tensor and the conformal gravity theories that include it. As we have noted above, the conformal invariance admitted by the Lanczos tensor, is basically induced by the conformal invariance of the Weyl tensor. Let us however emphasis that the conformal symmetry in the Weyl tensor, and hence in the Lanczos tensor, is admitted independently of the considered theory of gravity, which itself may or may not be conformally invariant. Let us recall that the Lanczos and Weyl tensors are basic ingredients in any metric theory of gravity (formulated in a 4-dimensional spacetime), for example GR. Nevertheless, GR is not a conformal theory.

A recently discovered conformal metric theory of gravity \cite{IJGMMP}, wherein the theory of Lanczos tensor is embraced naturally and solution (\ref{eq:sch}) is also admitted, might be worth mentioning. It is a simple unified theory of gravitation and electrodynamics formulated in a Ricci-flat spacetime, wherein the Bianchi identities play the role of the field equations  and electrodynamics appear through the symmetries of the spacetime. The theory is based primarily on
 the principles of equivalence and Mach, augmented with a novel insight that the field tensors in a geometric theory of gravitation and electromagnetism must be trace-free. Since these long-range interactions are mediated by virtual exchange of massless particles whose mass is expected to be related to the trace of the field tensors, the Riemann tensor (like the  electromagnetic Faraday tensor) must be trace-free, hence reducing to the Weyl tensor.

The vanishing of the Ricci tensor then appears as an initial/boundary condition in the theory. Hence, the successes of the classical GR in terms of the `source-free' Schwarzschild and Kerr etc. solutions, are readily embraced in this theory. Let us however emphasize that they do not represent empty spacetime in this theory. Rather they have a non-vanishing energy-momentum-angular momentum tensor of the gravitational field as their source. In fact, there is no `empty' spacetime in the theory and the energy, momentum and angular momentum of the material plus the gravitational fields emerge from the geometry itself. 
It has been shown \cite{IJMPD} that this conformal theory not only explains observations at all scales and solves the much hyped `Hubble tension', but also averts some long-standing problems of the standard cosmology, for instance, the problems related with the cosmological constant, the horizon, the flatness, the Big Bang singularity, the age of the universe and the non-conservation of energy. This is done without requiring any dark matter, dark energy or inflation.

Another conformal metric theory of gravity, which explains the galactic rotation curves (alternatively, not through the Lanczos tensor), has been proposed by Mannheim and coworkers \cite{Mannheim-Kazanas}. It is a fourth-order derivative theory derived from a conformal action quadratic in the Weyl tensor.
For a localized source, the theory generates not just the Newton potential but a local linear potential, together with  two global potentials as well. As has been mentioned above, these extra potentials are instrumental in explaining the galactic rotation curves  without any dark matter.
Besides, the theory also explains  the supernovae Ia observations without dark matter or dark energy \cite{Mannheim}.

\renewcommand{\theequation}{A-\arabic{equation}}
\setcounter{equation}{0}  % reset counter 
\section*{Appendix: {\bf A Brief Account of Lanczos Tensor}}

The Lanczos tensor discovered by Cornelius Lanczos  in 1962 \cite{Lanczos}, which happens to exist in 4-dimensions in any Riemannian manifold with Lorentzian signature, can be best understood in terms of the electromagnetic analogy. In electrodynamics, the electromagnetic 4-potential $A_\mu$ generates the conformally invariant, trace-free Faraday tensor $F_{\mu\nu}$
\be
F_{\mu\nu} = A_{\mu;\nu} - A_{\nu;\mu}.   \label{eq:F}
\ee
In the same way the Lanczos tensor $L_{\mu\nu\sigma}$ generates differentially the conformally invariant, trace-free Weyl tensor $C_{\mu\nu\sigma\rho}$ through the Weyl-Lanczos equation
\be
C_{\mu\nu\sigma\rho} = L_{[\mu\nu][\sigma;\rho]} +  L_{[\sigma\rho][\mu;\nu]} -  {*L*}_{[\mu\nu][\sigma;\rho]} -  {*L*}_{[\sigma\rho][\mu;\nu]},   \label{eq:Weyl-Lanczos}
\ee
if it satisfies the following two symmetries
\begin{subequations}
\label{eq:constraints}
\begin{align}
& L_{\mu\nu\sigma}=-L_{\nu\mu\sigma},\label{eq:syma}\\
& L_{\mu\nu\sigma}+L_{\nu\sigma\mu}+ L_{\sigma\mu\nu}=0. \label{eq:symb}
\end{align}
\end{subequations}
The double dual appearing in equation (\ref{eq:Weyl-Lanczos}) is defined by
${*M*}_{\alpha\beta\mu\nu}=\frac{1}{4}e_{\alpha\beta\rho\sigma}e_{\mu\nu\tau\delta} M^{\rho\sigma\tau\delta}$ and the square brackets [] enclosing indices denote skew-symmetrization: $N_{[\mu\nu]}\equiv \frac{1}{2!}(N_{\mu\nu}-N_{\nu\mu})$.
Augmenting the gravitation-electromagnetic correspondence further, the tensor $L_{\mu\nu\sigma}$ satisfies a homogeneous wave equation
\be
L_{\mu\nu\sigma;\kappa}^{~~~~~~~~;\kappa} =0, \label{eq:waveL}
\ee
in Lanczos gauge (\ref{eq:gauge}) in a Ricci-flat spacetime \cite{review, Edgar}, in the same way as  the vector $A_\mu$ satisfies the homogeneous wave equation
\be
A_{\mu;\kappa}^{~~~~;\kappa}  =0.\label{eq:Max}
\ee
 in Lorenz gauge ($A^\kappa_{~~;\kappa}=0$) in a Ricci-flat  spacetime \cite{Wald, IJGMMP}. 
Thus the Lanczos and Weyl tensors appear as the gravitational analogues of respectively the electromagnetic 4-potential and the Faraday tensor. This inspires to call the Lanczos tensor as the potential of the Weyl tensor.

By virtue of the symmetries (\ref{eq:syma}) and (\ref{eq:symb}), the tensor $L_{\mu\nu\sigma}$ possesses 20  independent components, while the Weyl tensor $C_{\mu\nu\sigma\rho}$ has only 10  independent components. 
 Lanczos considered the following additional symmetries
\begin{subequations}
\label{eq:gauge}
\begin{align}
 &  L^{~~\kappa}_{\mu~~\kappa}=0,         \label{eq:conditiona} \\
 & L^{~~~\kappa}_{\mu\nu~~;\kappa}=0,         \label{eq:conditionb}
\end{align}
\end{subequations}
as two gauge conditions in order to reduce the surplus degrees of freedom present in $L_{\mu\nu\sigma}$. 
The condition (\ref{eq:conditiona}) abolishes the degeneracy in $L_{\mu\nu\sigma}$ appearing through the gauge transformation
\be
\bar{L}_{\mu\nu\sigma} = L_{\mu\nu\sigma}+ g_{\nu\sigma}X_\mu - g_{\mu\sigma}X_\nu,   \label{eq:gaugeU}
\ee
which leaves  equation (\ref{eq:Weyl-Lanczos}) invariant for an arbitrary vector field $X_\alpha$. 
Thus the algebraic gauge condition (\ref{eq:conditiona}), taken together with (\ref{eq:syma}) and (\ref{eq:symb}), reduces
the number of independent components of $L_{\mu\nu\sigma}$ to 16.
Let us note that the differential gauge condition (\ref{eq:conditionb})
does not further reduce the degrees of freedom of $L_{\mu\nu\sigma}$, contrary to the wide-spread misunderstanding in the literature \cite{CQG,JMP}.
Thus ample degeneracy remains there in the values of the tensor $L_{\mu\nu\sigma}$ even after the Lanczos gauge conditions (\ref{eq:conditiona}) and (\ref{eq:conditionb}) are applied, as has been demonstrated in Ref. \cite{CQG}.
The cause of this degeneracy in $L_{\mu\nu\sigma}$ is the redundant degrees of freedom of the tensor, and it is not an issue with the gauge conditions. That is, the conditions  (\ref{eq:conditiona}) and (\ref{eq:conditionb}) alone, taken together with (\ref{eq:syma}) and  (\ref{eq:symb}), are not enough to determine a unique value of $L_{\mu\nu\sigma}$ in a given spacetime. In order to fix this arbitrariness, one needs additional conditions/assumptions of plausibility to be imposed on $L_{\mu\nu\sigma}$, which are still missing \cite{EPJC}.

The Riemann and Weyl tensors are the fundamental ingredients in a metric theory of gravity, which are supposed to provide a deeper understanding of the geometric character of the theory. Since the Lanczos tensor generates the Weyl tensor, it becomes even more fundamental and important. 
Here, we would like to emphasize that the above-sketched exposition on the Lanczos tensor is valid in any metric theory of gravity including GR, since it does not require the field equations of any particular theory of gravitation for its sustenance. Thus the Lanczos tensor emerges as an inherent structural element of any metric theory of gravity formulated in a 4-dimensional pseudo Riemannian spacetime.
However, this remarkable discovery, albeit its novelty and importance, has remained an obscure backwater to the mainstream relativists and cosmologists, even now - some sixty years after Lanczos first introduced it. 
The main reason for this obscurity is the absence of the physical meaning of the tensor, which has belittled an important discovery to a mere mathematical curiosity. It has not become possible so far to ascertain what the tensor represents physically, although some attempts have been made in this direction. 

For instance, Lanczos himself has showed an intimate relation of his tensor to Dirac's equation describing an electron with spin \cite{Lanczos}. 
In another study \cite{N-R}, the Lanczos tensor in Jordan's formulation of gravity leads to a model in which gravity and electroweak interactions are described in a unique framework.
In a recent study \cite{CQG}, it has been shown that the Lanczos tensor has a deeply ingrained quantum character which may lead to a gateway to quantum physics in the presence of gravitation. 
Another study \cite{EPJC} has discovered that the tensor can be portrayed as a relativistic, tensorial analogue of the Newtonian gravitational force. This claim has been corroborated by the matching of the tensor on the surface of a gravitating body. The claim is further supported by the construction of an energy momentum tensor from the Lanczos tensor, which represents  the energy-momentum of the gravitational waves  in a linearized theory.
Interestingly, the matching of the Lanczos tensor at the boundary of a collapsing mass, indicates towards a quantum signature of the tensor that can avoid the singularity \cite{EPJC}, which appears consistent with other studies \cite{CQG}.

These are only a few, but interesting, physical aspects  of the Lanczos tensor field which have only just started to reveal. Nonetheless, they open up new research avenues with rich prospects. However, there are various other surprises and insights, that need to be unearthed, and many questions that need to be addressed. For instance, how to find a general algorithm to find the Lanczos tensor given the metric tensor in any general spacetime? Further, as the Lanczos tensor appears to be the potential of the Weyl field and the latter can be written in terms of the Christoffel symbols, then how is the  Lanczos tensor related with the Christoffel symbols?

\end{document}